# Insights into Li$^+$, Na$^+$ and K$^+$ Intercalation in Lepidocrocite-type Layered TiO$_2$ Structures

*Kyle G. Reeves, *,[†] Jiwei Ma,[†] Mika Fukunishi,[‡] Mathieu Salanne,[†,⊥] Shinichi Komaba[‡] and Damien Dambournet*,[†,⊥]*

[†] Sorbonne Université, CNRS, Physico-chimie des électrolytes et nano-systèmes interfaciaux, PHENIX, F-75005 Paris, France

[‡]Department of Applied Chemistry, Tokyo University of Science, 1-3 Kagurazaka, Shinjuku, Tokyo 162-8601, Japan

[⊥]Réseau sur le Stockage Electrochimique de l'Energie (RS2E), FR CNRS 3459, 80039 Amiens, France

**Corresponding Authors**

* E-mail: kyle.reeves@sorbonne-universite.fr, damien.dambournet@sorbonne-universite.fr

# Abstract

A lamellar lepidocrocite-type titanate structure with ~25% $Ti^{4+}$ vacancies was recently synthesized, and it showed potential for use as an electrode in rechargeable lithium-ion batteries. In addition to lithium, we explore this material's ability to accommodate other monovalent ions with greater natural abundance (e.g. sodium and potassium) in order to develop lower-cost alternatives to lithium-ion batteries constructed from more widely available elements. Galvanostatic discharge/charge curves for the lepidocrocite material indicate that increasing the ionic radius of the monovalent ion results in a deteriorating performance of the electrode. Using first-principles electronic structure calculations, we identify the relaxed geometries of the structure for various positions of the ion in the structure. We then use these geometries to compute the energy of formations. Additionally, we determine that all ions are favorable in the structure, but interlayer positions are preferred compared to vacancy positions. We also conclude that the exchange between the interlayer and vacancy positions is a process which involves the interaction between interlayer water and surface hydroxyl groups next to the titanate layer. We observe a cooperative effect between structural water and OH groups to assist alkali-ions to move from the interlayer to the vacancy site. Thus, the as-synthesized lepidocrocite serves as a prototypical structure to investigate both the migration mechanism of ions within a confined space along with the interaction between water molecules and the titanate framework.

**Introduction**

Rechargeable lithium-ion batteries (LIBs) have become the norm in many modern-day electronic devices. With a recent interest in a greater production of battery-powered vehicles and a growing demand for batteries in electronic devices, battery production can be expected to increase dramatically in the coming decades.[1] Currently, these technologies focus on lithium-ion intercalation between positive and negative electrodes via a lithium-ion conducting electrolyte.[2,3] Attention was initially drawn to lithium for applications to rechargeable batteries because it leads to a high energy density and theoretical capacity.[4,5] Additionally, its small size makes intercalation in the electrode materials more feasible.

Despite the attractiveness of lithium, however, there remain challenges with its use. Notably, lithium is a rare element in the Earth's crust at ~20ppm.[6] Lithium is also non-uniformly distributed over the surface of the planet which could play a role in its availability for the production of LIBs as the demand for its use increases. Moreover, the extraction of lithium from natural sources such as hard rock, brine lakes or salt water is challenging and expensive.[7] Thus, future generations of rechargeable batteries would ideally take advantage of elements that are more abundant, more evenly distributed over the surface of the earth, and more easily processed.

Sodium and potassium, two other monovalent ions, are therefore promising candidates as ions to be used in place of lithium. The abundance of sodium and potassium ions is much greater than that of lithium, ranking 6$^{th}$ and 7$^{th}$ amongst all elements in the Earth's crust at ~23,000ppm and ~15,000ppm respectively. Additionally, the atomic weight and ionic size of each ion remain small enough to intercalate into electrode materials. The ionic radius of lithium, sodium and potassium ions grows from 0.76Å to 1.02Å and finally 1.38Å respectively. These differences in the atomic radii have shown that the performance of these materials can vary noticeably.[6,8–13]

An additional motivation for this research is the discovery of new electrode materials that can potentially incorporate a range of cations such as the ones mentioned before. In the first commercialized LIBs, graphite was used as the negative electrode. Today, rechargeable batteries continue to use carbon-based negative electrodes, but Ti-based materials are becoming increasingly common to study.[14–18] This shift is due in part to the processing necessary for carbon-based negative electrodes as well as safety concerns at the interface with the electrolyte once the battery is assembled.[19] A lamellar structure, such as that in graphite, has demonstrated the ability to accommodate a range of ions.[11] As a result of these factors, we, therefore, propose to investigate a Ti-based lamellar structure as a potential negative electrode material. We consider the lepidocrocite-type layering of $TiO_2$ which creates a lamellar structure stabilized by structural water. The role played by structural water during the intercalation process has been an intriguing question in different contexts. In a range of systems, it has been shown to influence the diffusion of ions, enhance structural stability upon cycling, and modify the redox mechanism.[20–24] Thus, we also consider this structure because of the potentially important role that the interlayer water may play in the intercalation process. As future technologies explore these new materials, it still remains unknown how the intercalation mechanism varies with the choice of the ion. In this work, we explore the intercalation of ions into a layered lepidocrocite-type titanate ($TiO_2$) structure. With the lepidocrocite titanate having already been both successfully synthesized and characterized,[25] we extend the investigation of this titanate material to probe its performance specifically with other positively charged monovalent ions.

**Experimental Measurements**

Using sol-gel chemistry, we recently synthesized an x-ray amorphous compound whose local structure was described to be as the lepidocrocite.[25] The experimental material was calculated to have a stoichiometry of $Ti_{1.5}\square_{0.5}O_2(OH)_2 \cdot 0.55H_2O$ where $\square$ represents a $Ti^{4+}$ vacancy whose charge is compensated by the additional hydroxide in the structure. The intercalation of alkali-ions into the structure can be expressed based on the available crystallographic sites provide by the titanium vacancy and interlayer space such as:

$[Ti_{1.5}\square_{0.5}]^{4h}O_2(OH)_2 \cdot [(H_2O)_n]^{4i}, + x.X^+ + x.e^- \leftrightarrow [Ti_{1.5}\square_{0.5-x}X_x]^{4h}O_2(OH)_2 \cdot [(H_2O)_nX_m]^{4i}$

where 4h and 4i refer to Wyckoff sites.[26] The calculated theoretical capacity based on the $Ti^{4+}/Ti^{3+}$ redox couple is 270 mAh.g$^{-1}$. The intercalation properties with respect to lithium, sodium, and potassium were assessed using galvanostatic electrochemical experiments in non-aqueous alkali metal half cells. The electrode was made using the lepidocrocite as the active material (80 wt%), carbon (10 wt%) and sodium carboxymethyl cellulose (CMC) (10 wt%) as the binder.[27,28] Coin-type cells are assembled with a lepidocrocite electrode, a glass fiber as a separator, a counter alkali metal electrode and filled with an electrolyte. 1 mol dm$^{-3}$ $LiPF_6$ dissolved in a mixture of 1:1 vol% of ethylene carbonate and dimethyl carbonate (EC:DMC), 1 mol dm$^{-3}$ $NaPF_6$ dissolved in a mixture of 1:1 vol% of ethylene carbonate and diethyl carbonate (EC:DEC) or 1 mol dm$^{-3}$ potassium bis(fluoroslufonyl)imide (KFSI) dissolved in a mixture of 1:1 vol% of EC:DEC are used as an electrolyte, and lithium, sodium, or potassium foil is used as a counter electrode for testing a Li$^+$, Na$^+$, or K$^+$ intercalation, respectively. All cell fabrications were done in the glove box filled with Ar gas. The cells were then cycled against the corresponding metals under 25 mA g$^{-1}$ in the voltage range of 0.0-3.0 V for Li cells and 0.0-2.0 V for Na and K cells at room temperature (ca. 25°C). The first three discharge/charge curves are shown in Figure 1a-c with cycling data gathered in Figure 1d. Overall, we observed that the reversible capacity decreases as the size of carrier ions increases. In the case of lithium, the reversible capacity was 180 mAh g$^{-1}$ which corresponds to the intercalation of one lithium per formula unit. In the case of sodium, the charge/discharge curves in Figure 1b show a notably lower charge capacity at around 120 mAh g$^{-1}$ which gradually increases to 135 mAh g$^{-1}$ at the end of the third cycle. In the last case of the potassium intercalation, we observed a gradual activation of the electrode with a reversible capacity of only 50 mAh g$^{-1}$ obtained at the end of the third cycle (Figure 1c). This points to a limited access of K$^+$ in the lepidocrocite structure. Attempts to improve the storage capacity by tuning the cell components such as the electrolyte was not successful. This points to a structural incompatibility rather than a non-optimized setup.

Over long-term cycling (Figure 1d), we observed a capacity decay which is more important as the size of the ion increases. Repeated lithium intercalation/de-intercalation reactions proceed in a stable manner with a capacity of 165 mAh g$^{-1}$ (0.92 Li$^+$ per formula unit) after 45 cycles. In the case of sodium, we observe a turnover in the discharge capacity before gradually decreasing to 90 mAh g$^{-1}$ (0.50 Na$^+$ per formula unit). Potassium intercalation stabilizes to a capacity of around 37 mAh g$^{-1}$ confirming the poor ability (*i.e.* 0.20 K$^+$ per formula unit) of the lepidocrocite structure in accommodating such a large cation.

Given that the material is of amorphous nature, which precludes the use of conventional x-ray diffraction and that the experimental cycling with different sources of alkali metals show markedly different performance, we turned to first-principles calculations to identify possible reasons for such differences.

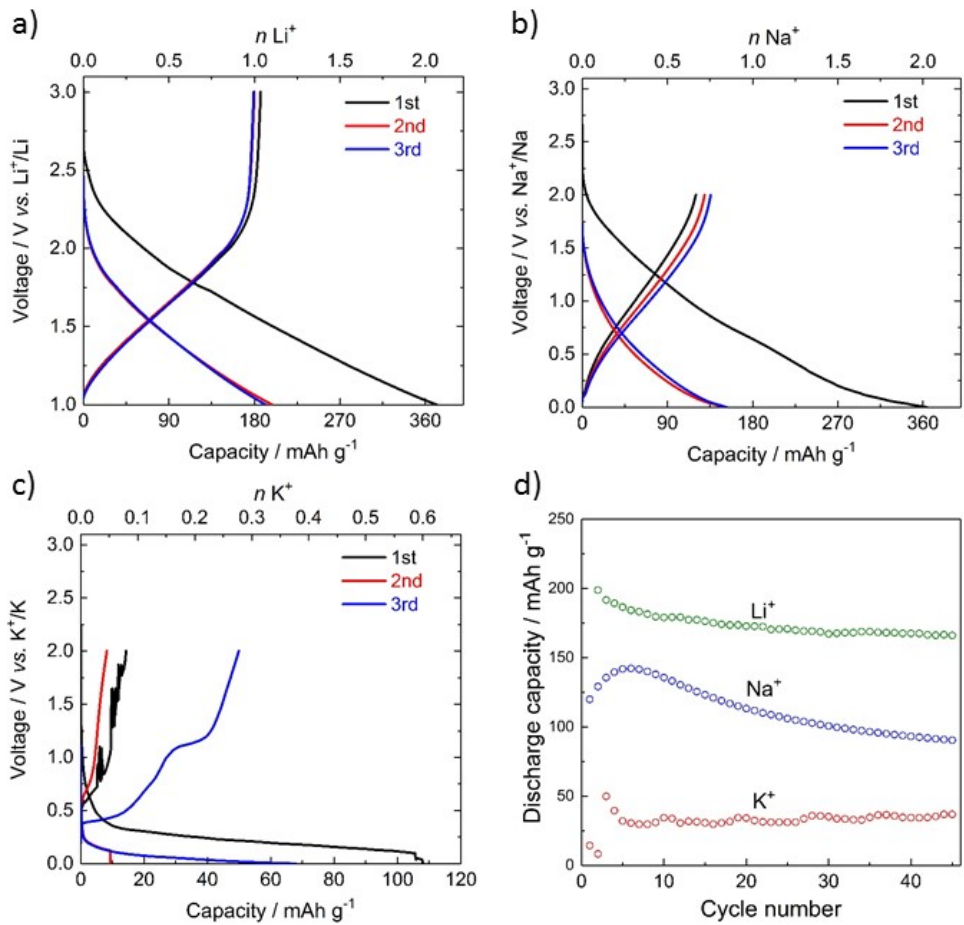

**Figure 1.** Galvanostatic discharge/charge curves obtained for cells cycled against metallic lithium (a), sodium (b) and potassium (c) under 25 mA g$^{-1}$. Cycling data (d).

Shown in Figure 2 and consistent with our discussion of the Wyckoff sites, we hypothesize that this material contains two host sites for the cations during intercalation: the interlayer position and the titanium vacancy. In the case of the interlayer position, the ions are incorporated into the material by being able to freely move within the plane of interlayer water, coordinated by surrounding water molecules. Ions in the Ti$^{4+}$ vacancy site are likely in a position where the transition metal would have been. In the vacancy, cations are coordinated by the surrounding oxygen atoms in the titanate layer.

Despite the structural complexity reported in the original material, we choose to simplify the system investigated computationally by proposing a simulation cell of $Ti_{63}\square_1O_{124}(OH)_4 \cdot 32H_2O$. These simplifications were made in order to limit the complexity of the calculations and focus on the contribution of each proposed host site. It can be seen that the proposed material contains only one vacancy for every 63 titanium atoms, a concentration of vacancies that is below that of the real structure. Additional electronic structure calculations suggest, however, that titanium vacancies in this structure are likely to be more stable when dispersed throughout the titanate layers. Therefore, the single titanate vacancy we simulate here in this model is likely an approximation that extends to the isolated vacancies that one would find in the real structure. The anionic environment of the titanium vacancies was generated via the substitution of hydroxyl groups, and the positions of each $OH^-$ was determined to be on the most under-coordinated oxygen atoms surrounding the Ti-vacancy. Similar in approach to the work by Grey and Wilson,[29] we determined that the optimal positions of the hydroxyl groups are the two oxygens at the surface of the titanate layer that are 1-fold coordinated and the two oxygens in the inner positions that are nearest to the vacancy. In this way, the four hydroxyl groups lead to charge balance.

To investigate the energetics of the system, we performed a series of first-principles density functional theory (DFT) calculations to determine the energy of formation for an ion in the structure. This quantity was calculated using the expression:

$$E_f = E[X + TiO_2] - (E[X] + E[TiO_2])$$

where the first term is the total energy of the lepidocrocite $TiO_2$ system with the ion, $X$, included in the system and the expression between the parentheses represents the sum of the neutral ion and lepidocrocite calculated independently. The first-principles electronic structure calculations were performed using the CP2K software via the Quickstep algorithm.[30,31] The generalized-gradient approximation PBE exchange-correlation functional was used.[32] For the basis set, we used DZVP-MOLOPT-SR-GTH[33] to construct the Kohn-Sham wavefunctions with a plane wave cutoff of 400 Rydberg. Goedecker-Teter-Hutter pseudopotentials[34] were used to describe core electrons, where Ti atoms were represented explicitly using $3s^23p^63d^24s^2$ electronic orbitals, and O atoms were represented using $2s^22p^4$ electronic orbitals. To simulate the system after the discharge process, each system is assumed to be charge neutral. Thus, the three cations were represented by explicitly representing the valence electrons with the following configurations: Li was represented using $1s^22s^1$, Na was represented using $2s^22p^63s^1$, and K was represented using $3s^23p^64s^1$.

Starting from a 4x4x1 lepidocrocite supercell, a system of $Ti_{63}\square_1O_{124}(OH)_4 \cdot 32H_2O$ was constructed to include a titanium vacancy, charge-compensating hydroxyl groups, and interlayer water molecules.[35] The spacing of the basal planes was fixed at 11.5Å, a value which is consistent with previous experimental work.[36] The calculation was performed with periodic boundary conditions and sampling only the Γ-point due to the large size of the simulation cell. For each calculation, we placed the ion into the structure and performed a geometry optimization keeping the ion position fixed. For the isolated lepidocrocite structure described by the second half of Equation 1, no further relaxation was performed.

By comparing the two host sites in the material, we can immediately see from Table 1 that all ions can be, from a thermodynamic perspective, favorably incorporated into the structure in either the vacancy or the interlayer positions all with negative formation energies expressed in meV/formula unit. Comparing between the cations, it is clear that lithium is the most favorable ion to be incorporated into the lepidocrocite structure in either of the two positions. For the vacancy position, we observe that the energy of formation increases, becoming less favorable as the size of the ion increases. This trend does not hold true, however, for the interlayer position, and although lithium shows the greatest stability, potassium seems to be the next most stable before sodium. The energy of formation for the ions in the interlayer are all within 20 meV.

**Table 1.** Energy of Formation (meV/formula unit) for ions in proposed host sites

|  | $Li^+$ | $Na^+$ | $K^+$ |
| --- | --- | --- | --- |
| Vacancy | -83.078 | -62.696 | -39.278 |
| Interlayer | -98.265 | -79.510 | -84.639 |

To investigate the environment of the ion in the interlayer, we compare the distance between the ion and the coordinating atoms as well as the number of coordinating molecules. In the case of the vacancy site, we take into consideration the distances between the ion and the nearby oxygens as it relates to the distortion that is generated in the titanate layer.

As we can see from Figure 2, the structures in the interlayer show an increasing bond length between the ion and the coordinating water oxygen. This trend is easily explained by each ion's ionic radius. As the size of the ion grows, the bond length naturally becomes longer. The ionic radius also explains the number of water molecules coordinating each ion. $Li^+$ and $Na^+$ both are both coordinated by three water molecules, but as $K^+$ has a larger ionic radius, it, therefore, is able to accommodate one additional water molecule. Unlike in the bulk, however, we note that the water molecules coordinate in a planar fashion. This has been observed in other lamellar structures[20], and we hypothesize that this coordination is due to the waters' role in both stabilizing the ion competing with its structural role in the lepidocrocite structure. On the other hand, all three ions were six-fold coordinated in the vacancy site (Figure 2d). The largest differences between these equilibrium structures are due to the amount of distortion that is introduced into the structure. As the ionic radius of the ion increases (*i.e.* $Li^+ < Na^+ < K^+$), the positions of the coordinating oxygens are pushed radially away from the center of the vacancy. The average distance from ion to the nearest neighbors increases from 2.20 Å in the case of Li$^+$ to 2.32 Å and 2.47 Å for the case of Na$^+$ and K$^+$ respectively.

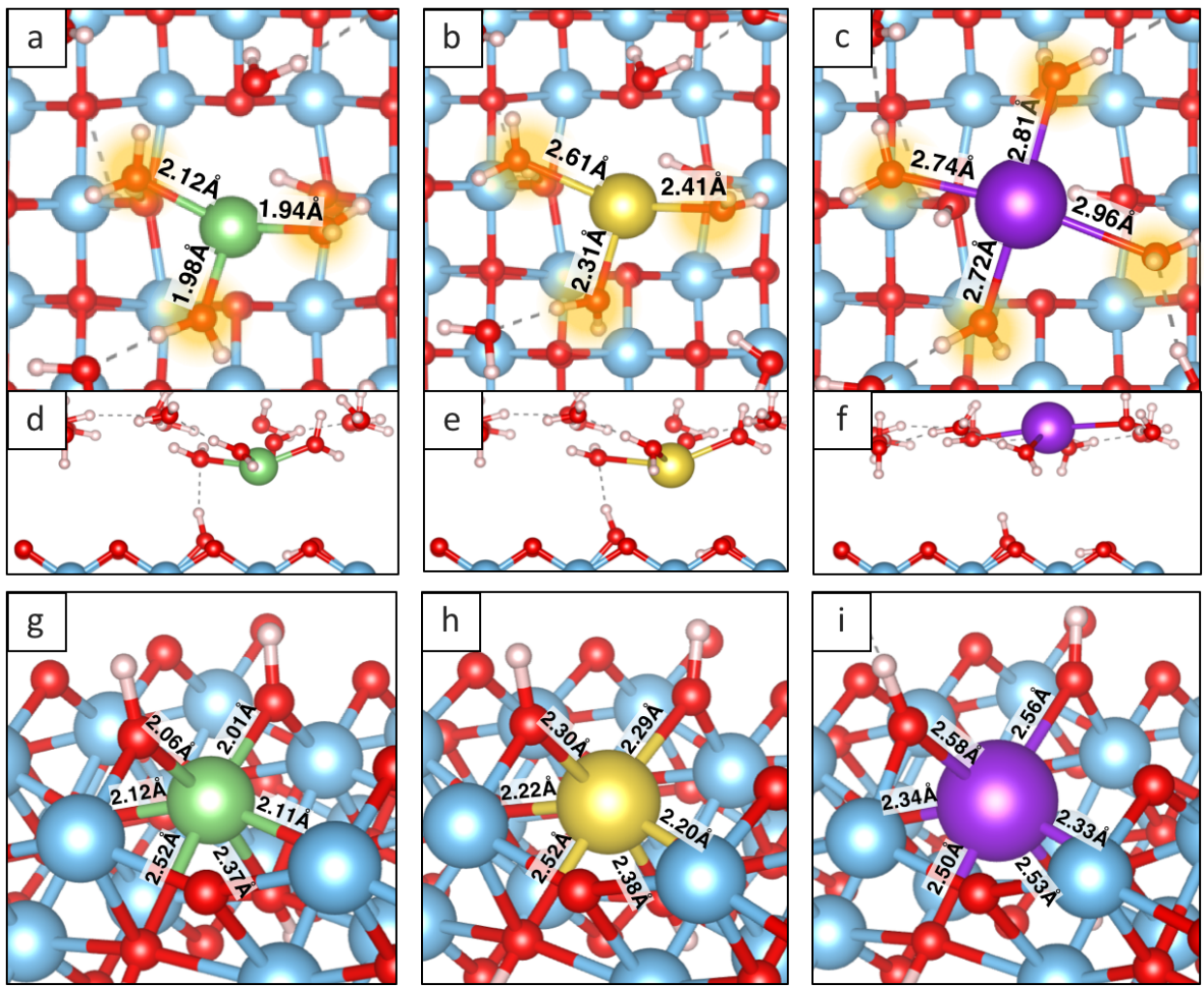

**Figure 2.** The equilibrium structures for the Li$^+$ (a, d, e), Na$^+$ (b, e, h), and K$^+$ (c, f, i). For each ion, the top panel shows a perspective that from above towards the Ti$^{4+}$ vacancy. The middle panel shows the position of the ion in the interlayer from an orthogonal direction. (a-f) represents interlayer structures. The equilibrium structure for all of the ions in the Ti$^{4+}$ vacancy (g, h, and i)

all show six-fold coordination yet with different average bond lengths (2.20 Å, 2.32 Å and 2.47 Å for $Li^+$, $Na^+$ and $K^+$ respectively).

This difference in ordering of ions between the two proposed sites suggests that there are different mechanisms of stabilizing the ions. We first explore the various relaxed structures calculated for each ion. In the case of the ions in the vacancy, we observe that structure surrounding the intercalated ion is displaced to a greater extent with the greater atomic radius. It is this distortion to the titanate structure upon ion insertion that contributes most significantly to the increase in the formation energy, leading to the order Li < Na < K. On the contrary, the stabilization of the Na and K are reversed in the interlayer due to their change in solvation structure.

Ion migration in these two-dimensional systems can be considered in two ways: within the plane of the interlayer and a perpendicular direction which connects the interlayer and the vacancy. Investigating diffusion of the ion within the interlayer is indeed feasible, but to converge the statistics to arrive at an accurate diffusion coefficient would require expensive molecular dynamics calculations. We instead choose to investigate the thermodynamics of system as the ion traverses in the perpendicular direction, which is expected to have much greater variation in total energy. Again, we investigate the energy of formation by placing the ion on a path that directly connects the two interlayer positions through the titanium vacancy (Figure 3, lower). In each position, we fix the position along the path and perform a geometry optimization that allows for the ion to move freely in the plane parallel to the titanate layer. In this way, we are able to construct an energetic curve that corresponds to a series of positions orthogonal to the titanate layer while allowing for some flexibility in the path that the ion may take between the two host sites.

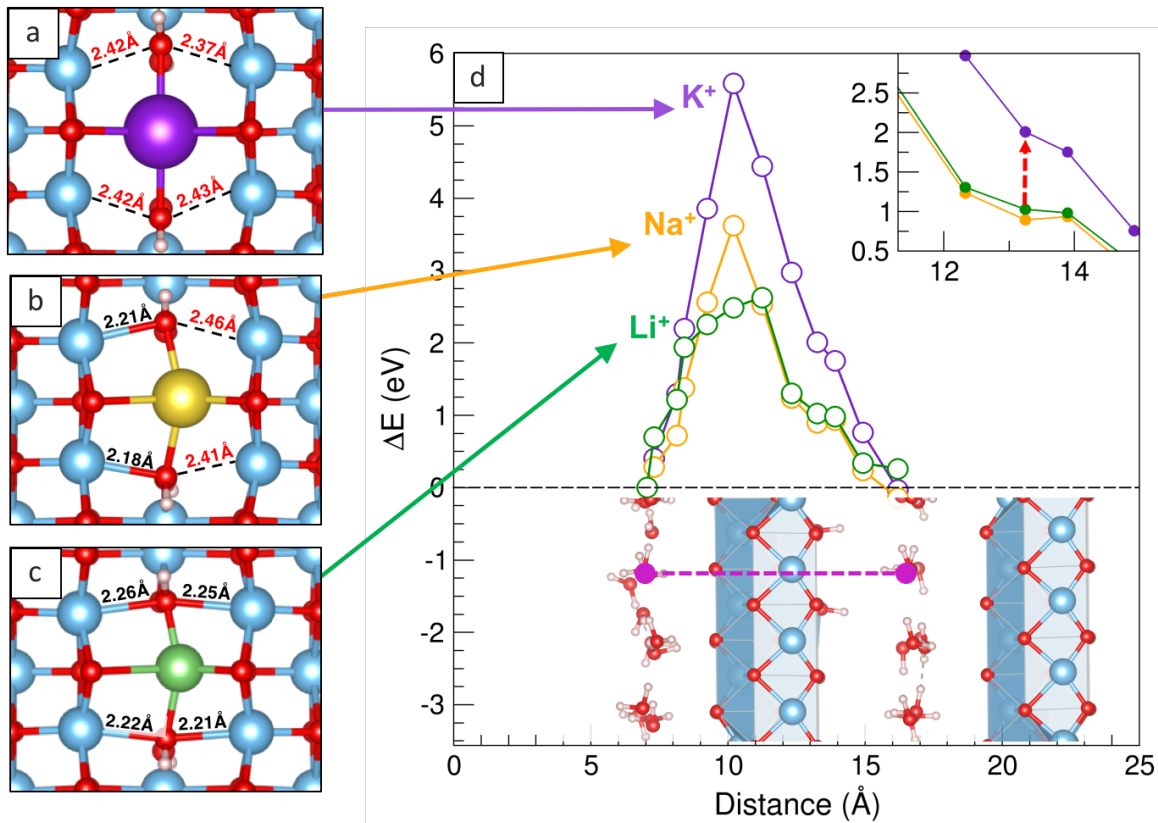

**Figure 3.** Structures of the lepidocrocite structure for the position of the greatest energy of formation. (a), (b) and (c) show the distortion of the lattice for $K^+$, $Na^+$, and $Li^+$ respectively, with distances marked in red indicating broken bonds. The energy of formation profile corresponding to a path between interlayers can be seen in (d).

We plot each curve as the difference in the energy of formation at a given point with respect to the lowest energy configuration. We note that although the difference in energies of formation is positive between positions on the ion path, the absolute energy of formation for all positions remains negative and therefore favorable. Consistent with Table 1, the lowest energy configuration was identified to be in the interlayer of the structure. The contribution from the Hartree energy (i.e. the electrostatic contribution of the electron density to the total energy) increases by 0.23eV/formula unit as the ion passes through titanate layer with the maximum increase corresponding to the peak shown in Figure 3. Unlike the vacancy position, ions in the interlayer are provided more space for water molecules to coordinate to the ion without confinement. This flexibility in the local environment to accommodate the ion is likely the reason for the interlayer being the preferred site of the lepidocrocite structure.

The maxima in the curve occur around 10.5Å in the simulation cell, which corresponds to the plane of oxygen atoms which are 4-fold coordinated to surrounding titanium atoms. Unlike the surface oxygen atoms which benefit from the flexibility of the structure due to the lower coordination, oxygen atoms are much more sensitive to being displaced. Thus, the changes that we observe in the energy profile at this location is due almost entirely to the distortion of the structure. On the left-hand side of Figure 3, we provide the Ti-O bond lengths associated with the oxygen atoms nearest to the interpenetrating cation. Those bond lengths that are indicated in red represent distances that are longer than those in the original, equilibrium structure without the ion and which are also greater than what is generally accepted as a typical Ti-O bond length.

In each case, as the ion passes through the titanate layer, the structure must distort in order to accommodate the cation. Between 10Å-12Å in the simulation cell, the ion moves through the structure into a region with a plane of oxygen atoms. At this point in the path, the ion is the closest to any other neighboring atoms than at any other point in the simulation.

We observe that the difference in the energy of formation between the sodium and the lithium ion between the interlayer and vacancy (Figure 3, inset) are comparable whereas the potassium ion shows a greater energy difference. It is the flexibility of the local environment of the adjacent hydroxyl groups at the titanate surface which allows for the titanate layer to accommodate the sodium ion and leads to a similar stability for lithium and sodium. This flexibility is not sufficient, however, to accommodate the larger potassium ion without imposing a stress on the surrounding structure. Hence, we conclude that the low electrochemical activity of the lepidocrocite with respect to potassium is due to the inability of the vacancy to accommodate such a large cation. Moreover, as potassium is solvated by a greater number of water molecules, we hypothesize that its intercalation may also be limited by the availability of structural water molecules to stabilize the ion.

In addition to the electronic interactions directly between the titanate structure and the ions, it is important to consider the role that solvent molecules will play in the interlayer and for the reversible insertion of the ions into the titanium vacancies. As we saw from the first section, the coordination of the ion stabilizes it in the interlayer, and so the process of destabilizing the ion must play a role in the transfer from one host site to another. Thus, the question remains if the water molecules that are coordinated to the ion in the interlayer play any role in the insertion into the vacancy.

To investigate the possible role of water in exchange of ions between the interlayer sites and the vacancy sites, we set up a series of calculations that were designed to better understand the role that interlayer water may play in ushering ions between the two proposed host sites. We begin with the ion in the interlayer of the system in its relaxed configuration. We then manually move the ion in the direction of the vacancy using the positions of the interlayer water from the previous step. The structure is again relaxed, keeping the position of the ion fixed. This iterative process of advancing the ion and relaxing the local water molecules allows us to bias the motion of the ion and observe how the local solvation environment responds to the perturbation. This process is not intended to prove any specific mechanism of ion insertion, but rather to generate insight into the role that coordinating water molecules may play in facilitating the insertion. Given the already large energetic penalty association with the distortion of the lattice for the potassium ion, our analysis focuses on the motion of the sodium ions. Several frames of the path for the sodium ion can be seen in Figure 4. As the ion passes from the interlayer to the vacancy, two of the three water molecules that were initially coordinated to the ion follow the motion of the sodium. As the ion approaches the vacancy, it reaches tetrahedral coordination (Figure 4b) with the two water molecules and the two hydroxyl groups at the surface. When the ion is inside of the vacancy, we see that the equilibrium geometry no longer includes the direct coordination of these two water molecules (Figure 4d). This simulation makes it clear that the movement of interlayer water, in an effort to continually stabilize the ion, plays an important role in the movement of ions between host sites. Moreover, we see that it is not just the water molecules that play a role, but also the hydroxyl groups facing the interlayer. It is therefore a concerted and cooperative effect that allows for the movement between sites in the structure. The motion of the hydroxyl groups is not as free as the water molecules, however, and their displacement in turn affects the forces felt by the surrounding titanate structure. Figure 4c shows a frame in the path of the ion where it is coordinated by the hydroxyl groups by equal distance on either side of the ion.

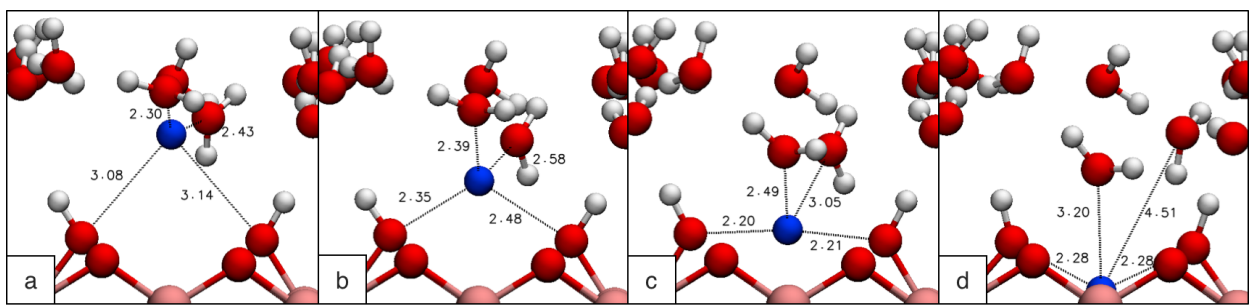

**Figure 4.** The path of a sodium ion as it moves from interlayer to the $Ti^{4+}$ vacancy with interatomic distance labeled. As the ion moves towards the vacancy, two water molecules (a) follow the ion until it is tetrahedrally coordinated (b). As the ion continues to move, it is symmetrically coordinated by the hydroxyl groups at the surface of the titanate layer (c) until ultimately the initially coordinated water molecules no longer strongly interact with the ion in the vacancy (d). Distances are labeled in angstroms.

During the ion migration process, the hydroxyl groups are displaced away from their equilibrium position (Figure 5a). We seek to identify the individual contributions of the interlayer water and the titanate distortion separately. Figure 5b shows the total energy of the system as the sodium ion moves from the interlayer to the vacancy. In solid blue, we show the total energy of the system with respect to the energy of the system in the interlayer when the coordinating water molecules do not accompany the sodium ion and the hydroxyl groups are allowed to move. We can compare this to the same path with the hydroxyl groups fixed in their equilibrium position (solid green line) and see that there is a dramatic increase in total energy of the system when the structure is no longer allowed to adjust to accommodate the size of the ion. It is clear that the flexibility of the framework is an important factor in stabilizing the movement of the ion. We estimate that the greatest deviation in the energy is at the position that bisects the distance between the two oxygens showing a difference in energy of ~6.5eV. The curve shown in red is the total energy of the system as both the coordinated waters follow the ion and the hydroxyl groups adjust. The combination of these two processes shows a significant reduction in the total energy of the system. Thus, this process may include an aspect of stabilization via the coordination of the ion while simultaneously introducing an energetic penalty associated with the ion driving a distortion in the lattice away from its equilibrium structure.

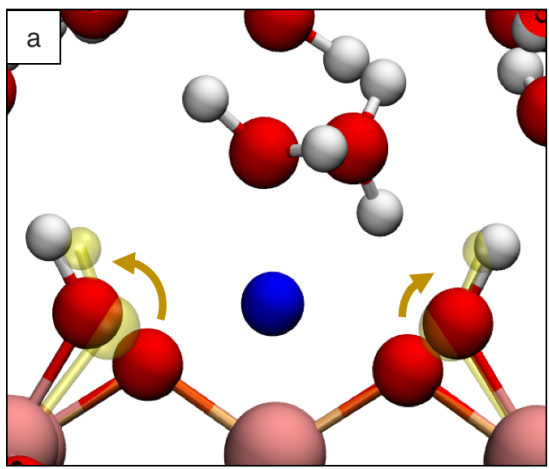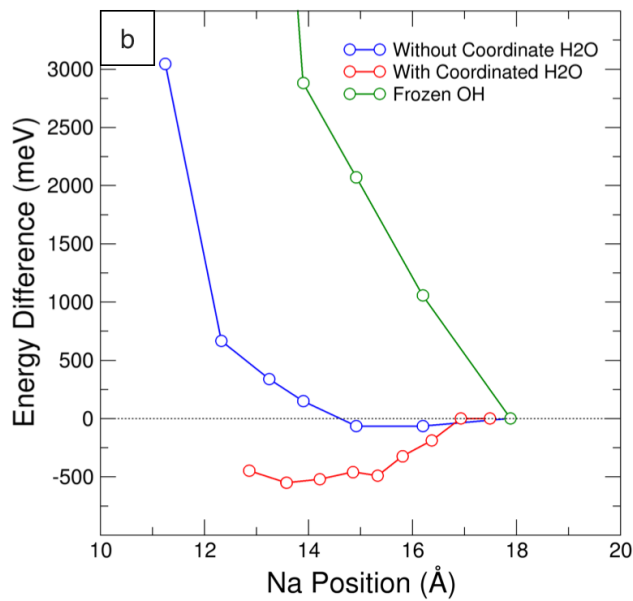

**Figure 5.** The motion of the ion towards the vacancy induces a pivoting of the hydroxyl group (a) away from the vacancy to accommodate the size of the ion. The equilibrium positions of the hydroxyl groups in the absence of the ion are shown in yellow. The total energy of the system as the ion moves from the interlayer to the vacancy (b) where the hydroxyl group positions are fixed and no water molecules accompany the ion (solid green), ions are not coordinated by water molecules, but hydroxyl groups are free to move (solid blue), and a system where water molecules move with the ion and hydroxyl groups are free to move (solid red). The energies in (b) are changes in total energy with respect to the equilibrium interlayer configuration and differ from the energy in Figure 3 which represents the energy of formation where each data point is in reference to its locally relaxed geometry.

**Conclusions**

The lepidocrocite-type titanate material has already shown promising performance as a negative electrode with lithium as the intercalating ion. The different performance of the ions we observed in our investigation suggests that the material can also serve the important role as a prototypical material to understand diffusion and transport of ions in lamellar structures and in particular those materials which contain structural water. Additionally, although the energy of formation theoretically predicts favorable intercalation into both the interlayer position and the vacancy for all ions, we conclude that the likelihood is extremely reduced as the size of the ion increases. For lithium, sodium and potassium ions, the interlayer is a preferred site due to the surrounding coordination from structural water present in the interlayer. Ions, when found in the titanate layer, are likely to be located in the $Ti^{4+}$ vacancies, however large ions such as potassium that introduce a significant amount of distortion to the lattice result in a greater energy penalty. Features such as $Ti^{4+}$ vacancies, which generate a greater amount of flexibility in the structure, were shown to be unable to accommodate the large $K^+$ ion without distortion. We additionally investigate the role that the interlayer water plays in the movement of ions between the interlayer and vacancy sites. By comparing the path of the ions passing from the interlayer to the vacancy through vacuum as compared to the movement of the water coordinated ions, we conclude that the total energy is significantly reduced by nearly 0.5eV when water and hydroxyl groups facilitate the motion of the ion towards the vacancy. The interlayer water, therefore, plays a dual role: a structural role in the lamellar structure as well as a role of stabilizing and accompanying ion motion in the interlayer. Other materials with structural water, especially those which may have surface hydroxyl groups may also exhibit a stabilization of intercalated ions.

**Acknowledgments**

The research leading to these results has received funding the cluster of excellence MATeriaux Interfaces Surfaces Environnement (MATISSE). We are grateful for the computing resources on CURIE (TGCC, French National HPC) obtained through Project No. A0010907684.